\documentclass{aa}
\usepackage{txfonts,makecell}
\usepackage{hhline}
\usepackage{hyperref}
\usepackage{cleveref}
\usepackage[alsoload=astronomy]{siunitx}

\crefname{equation}{eq.}{eqs.}

\newcounter{mysfig}
\renewcommand\themysfig{(\alph{mysfig})}
\makeatletter
\newcommand\Scaption[1]{%
\refstepcounter{mysfig}%
\vskip.5\abovecaptionskip
  \sbox\@tempboxa{\small\themysfig~#1}%
  \ifdim \wd\@tempboxa >\hsize
    \small\themysfig~#1\par
  \else
    \global \@minipagefalse
    \hb@xt@\hsize{\hfil\box\@tempboxa\hfil}%
  \fi
  \vskip\belowcaptionskip}
\makeatother
\newcommand{\diff}{\mathrm{d}}

\begin{document} 
\title{Strong evidence for an accelerating universe}
  
  \author{
  	Balakrishna S. Haridasu\inst{\ref{inst3}}
    \and Vladimir V. Lukovi\'{c}\inst{\ref{inst1},\ref{inst2}}
    \and Rocco D'Agostino\inst{\ref{inst1},\ref{inst2}}
    \and Nicola Vittorio\inst{\ref{inst1},\ref{inst2}}
  }
  
  \institute{Gran Sasso Science Institute (INFN), Viale Francesco Crispi 7, I-67100 L’Aquila, Italy\label{inst3} \and Dipartimento di Fisica, Universit\`{a} di Roma "Tor Vergata", Via della Ricerca Scientifica 1, I-00133, Roma, Italy\label{inst1}\and Sezione INFN, Universit\`{a} di Roma "Tor Vergata", Via della Ricerca Scientifica 1, I-00133, Roma, Italy\label{inst2}}

  \offprints{sandeep.haridasu@gssi.infn.it}

  \date{Received / Accepted }

  \abstract{A recent analysis of the Supernova Ia data claims a 'marginal' ($\sim3\sigma$) evidence for a cosmic acceleration. This result has been complemented with a non-accelerating $R_{h}=ct$ cosmology, which was presented as a valid alternative to the $\Lambda$CDM model. In this paper we use the same analysis to show that a non-marginal evidence for acceleration is actually found. We compare the standard Friedmann models to the $R_{h}=ct$ cosmology by complementing SN Ia data with the Baryon Acoustic Oscillations, Gamma Ray Bursts and Observational Hubble datasets. We also study the power-law model which is a functional generalisation of $R_{h}=ct$. We find that the evidence for late-time acceleration is beyond refutable at a 4.56$\sigma$ confidence level from SN Ia data alone, and at an even stronger confidence level ($5.38\sigma$) from our joint analysis. Also, the non-accelerating $R_{h}=ct$ model fails to statistically compare with the $\Lambda$CDM having a $\Delta(\text{AIC})\sim30$.}
 
\maketitle

\section{Introduction}
The very first evidence for an accelerated expansion of the universe was obtained using the SN Ia observations in \cite{Riess98} and \cite{Perlmutter99}, which has been further confirmed with the most recent supernova data \citep{Betoule14}. Other low redshift data such as the Observational Hubble parameter-OHD \citep{Jimenez02}, Baryon Acoustic Oscillations-BAO \citep{Eisenstein05}, also support an accelerating universe. As an independent observation, the Cosmic Microwave Background-CMB radiation has been in excellent concert with these results and has provided with the most stringent constraints on the cosmological models \citep{Ade16}. These observations have established the $\Lambda$CDM as the concordance model of cosmology and the late-time acceleration has been a well accepted phenomenon. 

However, \cite{Nielsen15} have used a modified statistical model for the analysis of the supernova JLA dataset \citep{Betoule14} and claimed that the evidence for the acceleration is marginal $(\lesssim 3\sigma)$. The modification has been done by assuming an intrinsic variation in the SN absolute magnitude and in the light curve (colour and stretch) corrections, which were modelled as Gaussian. More recently, \cite{Rubin16} have strongly criticised this approach as incomplete and suggested using redshift dependent ad hoc functions for these corrections, presenting the evidence for acceleration to be $\sim 4.2 \sigma$. In this paper we want to show that even with the less flexible modelling of \cite{Nielsen15} the evidence for acceleration is very strong. We also extend the joint analysis done in \cite{Lukovic16} with the inclusion of Gamma Ray Bursts-GRB dataset \citep{Wei10}. 

The marginal evidence for an accelerating universe quoted in \cite{Nielsen15} implies a scenario with very low dark matter and dark energy densities. As this scenario converges towards the Milne model, it has been complemented with the $R_{h}=ct$ cosmology \citep{Melia12c,Melia12a,Melia12b}, which features a non-accelerating linear expansion of the universe. This model essentially advocates that the Hubble sphere is same as the particle horizon of the universe \citep{Bikwa12}. Also, the $R_{h}=ct$ model has often been regarded as a Milne universe \citep{Mitra14,Bilicki12} and several physical issues against this interpretation have been raised \citep{Lewis13,Lewis12}. In fact, this model preserves the linear expansion by imposing the constraint on the total equation of state (EoS) ($\rho_{tot} + 3p_{tot}=0 $), without requiring $\rho_{tot} = p_{tot} = 0$. It is interesting to note that this model also coincides with the linear coasting models that are discussed in the context of modified gravity \citep{Gehlaut03,Dev00,Dev02}. In a recent work, \cite{Kumar16} presented a linear coasting model in $f(R)$ gravity, which is functionally equivalent to the $R_{h}=ct$ model. A linear expansion model can be generalised to a power-law model with an exponent $n$, which was brought up as an alternative to the standard model as it does not have the flatness and horizon problems \citep{Sethi05}. In addition, it has been shown in \cite{Dolgov97} that classical fields coupling to spacetime curvature can give rise to a back-reaction from singularities, which can change the nature of expansion from exponential to power-law. 
 
The $R_{h}=ct$ model has been tested time and again and contradictory conclusions have been presented. \cite{Bilicki12} have pointed out some of the problems in this model using the cosmic chronometers data from \cite{Moresco12a,Moresco12b} and radial BAO measurements, along with several model-independent diagnostics, to show that $R_{h}=ct$ is not a viable model. In \cite{Melia14a} several claims against $R_{h}=ct$ were refuted, and a list of works favouring $R_{h}=ct$ over $\Lambda$CDM was compiled in \cite{Melia16}. Power-law cosmologies with $n \geq 1$ have been explored against data in several works such as \cite{Gehlaut03, Dev08, Sethi05, Zhu08, Shafer15,Rani15,Dolgov14}, finding $n \sim 1.5$ consistently. In a more recent work \citep{Shafer15}, power-law and $R_{h}=ct$ models were tested with SN Ia and BAO datasets and were found to be highly disfavoured against $\Lambda$CDM. As most of these works have used older data for their analyses, we believe this is a good occasion to revise the constraints using more recent data and hence statistically verify the viability of these models against $\Lambda$CDM. 

The present paper is structured as follows. A brief introduction to the models is given in \Cref{sec:mod}. We describe the data and method used for our joint analysis in \Cref{sec:data}. Our results and discussion are given in \Cref{sec:ana}.

\section{Models}
\label{sec:mod}
In this section we briefly describe the standard $\Lambda$CDM model, power-law and $R_h=ct$ cosmologies, which we test to asses the late-time acceleration. The dominant components of the $\Lambda$CDM model at late times are cold dark matter (CDM), treated as dust, and dark energy (DE) fluid with an EoS parameter $w=-1$. 
The corresponding Friedmann equation is given by,
 \begin{equation}
\label{eqn:HUE}
H(z)^{2} = {H_0}^2\left( \Omega_{m}(1+z)^3  + \Omega_{k}(1+z)^2 +\Omega_{\Lambda}(1+z)^{3(1+w)}\right),
 \end{equation}
where $H_{0}$ is the present expansion rate, while $\Omega_{m}$, $ \Omega_\Lambda$ and $ \Omega_k $  are the dimensionless density parameters. 
For the flat $\Lambda$CDM model, $\Omega_{m} =1-\Omega_{\Lambda} $. We test the extensions of $\Lambda$CDM model, namely the $k\Lambda$CDM model with the constraint $\Omega_{m} =1-\Omega_{\Lambda}-\Omega_{k} $, and the flat $w$CDM model with $w$ as a free parameter.
 The second Friedmann equation, $ \ddot{a}/a=- 4\pi/3 G\sum_i\rho_i(1+3w_i)$, gives us insight into the necessary conditions to be satisfied for assessing the dynamics of expansion rate. The criteria for acceleration can be derived as: $\Omega_{m} \leq \Omega_{\Lambda}/2$ for $k\Lambda$CDM and $w \leq -1/(3 \Omega_{\Lambda})$ for $w$CDM. We can asses the evidence for acceleration by estimating the confidence level with which the criteria are satisfied. 
 
In a flat, power-law cosmological model the scale factor evolves in time as $a(t) \propto t^{n}$, with the Hubble equation $H(z) = H_{0}(1+z)^{1/n}$. Here, $n>1$ implies an accelerated scenario. Although motivated physically with a total EoS parameter $w_{tot}=-1/3$, the flat $R_{h}=ct$ model coincides with the power-law model for $n=1$. 
It is worthwhile noting that \Cref{eqn:HUE} reduces to the functional form of a power-law model for selected parameter values:   $ \Omega_{m} = 0 $, $\Omega_{\Lambda}=1$ and  
\begin{equation}
\label{eqn:wn}
w = \frac{2-3n}{3n}.
\end{equation}

The luminosity distance for all these models with their corresponding $H(z)$ is written as,
 \begin{equation}
 \label{eqn:luD}
D_{L}(z) \equiv \left\{
        \begin{aligned}
          & (1+z) c \left( \int_{0}^{z} \frac{\diff\xi}{H(\xi)}\right)  &&\text{for}\,\,\Omega_{k}=0 \\
          & \frac{(1+z) c}{H_{0}\sqrt{-\Omega_{k}}} \sin\left( \sqrt{-\Omega_{k}} H_{0} \int_{0}^{z} \frac{\diff\xi}{H(\xi)}\right)  &&\text{for}\,\,\Omega_{k}\neq0  \\
        \end{aligned}\right.
 \end{equation} 
 The theoretical distance modulus is defined as $\mu_{th} =  5\log[D_{L}(Mpc)] + 25$. The angular diameter distance is $D_{A}(z)=D_{L}(z)/(1+z)^{2}$, which is used in the modelling of BAO data.
 
\section{Data and Method}
\label{sec:data}
We test the models described in the \Cref{sec:mod} against data in the redshift range $0<z\lesssim 8$. We use the observables SN Ia, BAO, OHD and GRB that are uncorrelated. We perform a joint analysis using all the datasets together by defining a combined likelihood function. We keep the description of the data to a minimum as we refer to \cite{Lukovic16} for most of it. 

We use the JLA dataset \citep{Betoule14} consisting of 740 SN, which already provides an empirical correction to the absolute magnitude,
\begin{equation}
M^{corr}_{B} = M_{B} -\alpha s + \beta c,
\end{equation}
In the statistical method we implement \citep{Nielsen15}, the stretch $s$, colour $c$ corrections and the absolute magnitude $M_{B}$ are all considered random Gaussian variables without any redshift dependence. Such an assumption does not account for the selection effects in 's' and 'c' corrections. The JLA dataset has been corrected for the selection bias only in the apparent magnitude \citep{Betoule14}, which is why the correction in 's' and 'c' have to be explicitly included when they are modelled as distributions. As anticipated in the introduction, we use the less flexible modelling of \cite{Nielsen15} to show that even in this case the evidence for acceleration is very strong. Different methods for treating the selection bias in the SN data and their short comings have been discussed in \cite{Kessler17}. It is of high importance to study these effects, which we shall address in a forthcoming paper. The SN Ia likelihood $ {\cal L}_{\rm SN} $ used here is described in \cite{Nielsen15, Shariff15, Lukovic16}. 

The BAO data is available for the compound observable $D_{V}$ defined in \cite{Eisenstein05},
\begin{equation}
D_{V}(z) = \left[(1+z)^2 D_{A}^{2}(z)\frac{c z}{H(z)}\right]^{1/3} \ ,
\end{equation}
It is important to note that the observable $D_{V}$ is usually presented as a ratio with $r_{d}$, the sound horizon at the drag epoch. For the purpose of model selection we fit $r_{d}$ as a free parameter instead of using the standard fit function for the drag epoch $z_{d}$ \citep{Eisenstein98}, as it is not very straight forward to use it for the power law cosmologies. A similar approach was also implemented in \cite{Shafer15}. Hence, the parameters $H_{0}$ and $r_{d}$ are now degenerate, and BAO data by itself is only able to constraint the combination $r_{d} \times H_{0}$ and $\Omega_{m}$.  
To avoid correlations among different BAO data points, we use only six measurements taken from \cite{Beutler11, Anderson14, Ross15, Delubac15, Ribera14} also summarised in Table 1 of \cite{Lukovic16}. 
A simple likelihood for the uncorrelated data is then implemented as,
\begin{equation}
 {\cal L}_{\rm BAO}\propto\exp\left[-\dfrac{1}{2} \sum_{i=1}^{6} \left(\frac{ r_{d}/D_{V}^{i}-r_{d}/D_{V}(z_i)}{\sigma_{r_{d}/D_{V}}^{i}}\right)^2\right].
 \end{equation}

The measurements of the expansion rate have been estimated using the differential age (DA) method suggested in \cite{Jimenez02}, which considers pairs of passively evolving red galaxies at similar redshifts to obtain $\diff z/\diff t$. 
We use a compilation of 30 uncorrelated DA points taken from \cite{Simon05, Stern10, Moresco12b, Moresco16a, Moresco15, Zhang14} obtained using BC03 models.
We implement a simple likelihood function assuming all the data are uncorrelated. 
\begin{equation}
{\cal L}_{\rm OHD}\propto\exp\left[-\dfrac{1}{2}\sum_{i=1}^{30} \left(\frac{ H_i-H(z_i)}{\sigma_{H_i}}\right)^2\right].
\end{equation}

Finally, we use the GRB dataset comprising of 109 observations compiled with the well known Amati relation \citep{Amati02,Amati08,Amati09}. The dataset has 50 GRBs at $z<1.4$ and 59 GRBs at $z>1.4$, in a total range of $0.1<z<8.1$. The dataset is given in tables I and II of \cite{Wei10}. The distance modulus $\mu_{GRB}$ and the corresponding standard deviation can be defined as,
\begin{eqnarray}
\mu_{GRB} &=& \frac{5}{2}\left(\log_{10}\left[ \frac{(1+z)}{4\pi}\left(\frac{E_{p,i}}{300 \si{\kilo\electronvolt}}\right)^{b}\frac{S_{bolo}^{-1}}{100 \parsec^{2}} \right] + \lambda \right) \\
\sigma_{\mu_{GRB}} &=& \left(\frac{5}{2\log(10)}\right)^{2}\left[\left(\frac{b \sigma_{E_{p,i}}}{E_{p,i}}\right)^{2} + \left(\frac{\sigma_{S_{bolo}}}{S_{bolo}}\right)^{2}+\sigma_{sys}^{2}\right]. 
\end{eqnarray}
We adopt $\sigma_{sys} = 0.7571$, following the model independent calibration done in \cite{Feng16}. 
The likelihood for the GRB is defined as, 
\begin{equation}
{\cal L}_{\rm GRB}\propto\exp\left[-\dfrac{1}{2}\sum_{i=1}^{109} \left(\frac{\mu_{GRB}^{i} - \mu_{th}^{i} }{\sigma_{\mu_{GRB}}^{i}}\right)^2\right].
\end{equation}

The joint likelihood for these four independent observables is given as $\cal{L}_{\rm tot}=\cal{L}_{\rm SN}\cal{L}_{\rm OHD}\cal{L}_{\rm BAO}\cal{L}_{\rm GRB}$. We use the two most common criteria for model comparison in cosmology, namely the Akaike Information criteria (AIC) \citep{Akaike74} and the Bayesian Information criteria (BIC) \citep{Schwarz78}.
The AIC and BIC values for large number of measurements are defined as,
\begin{align}
\text{AIC} &= -2\log{\cal L}^{max} + 2 N_p , \\
\text{BIC} &= -2\log{\cal L}^{max} + N_p\log(N_{data}),
\end{align} 
where, $N_p$ and $N_{data}$ are the number of parameters and data points, respectively. $\Delta$(AIC) = AIC-AIC$_{ref}$ 
criterion takes into account the number of parameters to estimate the amount of information lost in one model when compared to a reference model, in our case $\Lambda$CDM. We define the $\Delta$(BIC) similar to $\Delta$(AIC). A negative value of the $\Delta$(AIC) or $\Delta$(BIC) indicates that the model in comparison performs better than the reference model. 

\section{Results and discussion}
\label{sec:ana}
In this Section we present the results obtained from our joint analysis for the models and data given in the earlier Sections. We first present our assessment for the current accelerated state of the universe and then comment on the model comparison using the AIC and BIC statistics.

\begin{figure}[h]
\includegraphics[width=0.45\textwidth]{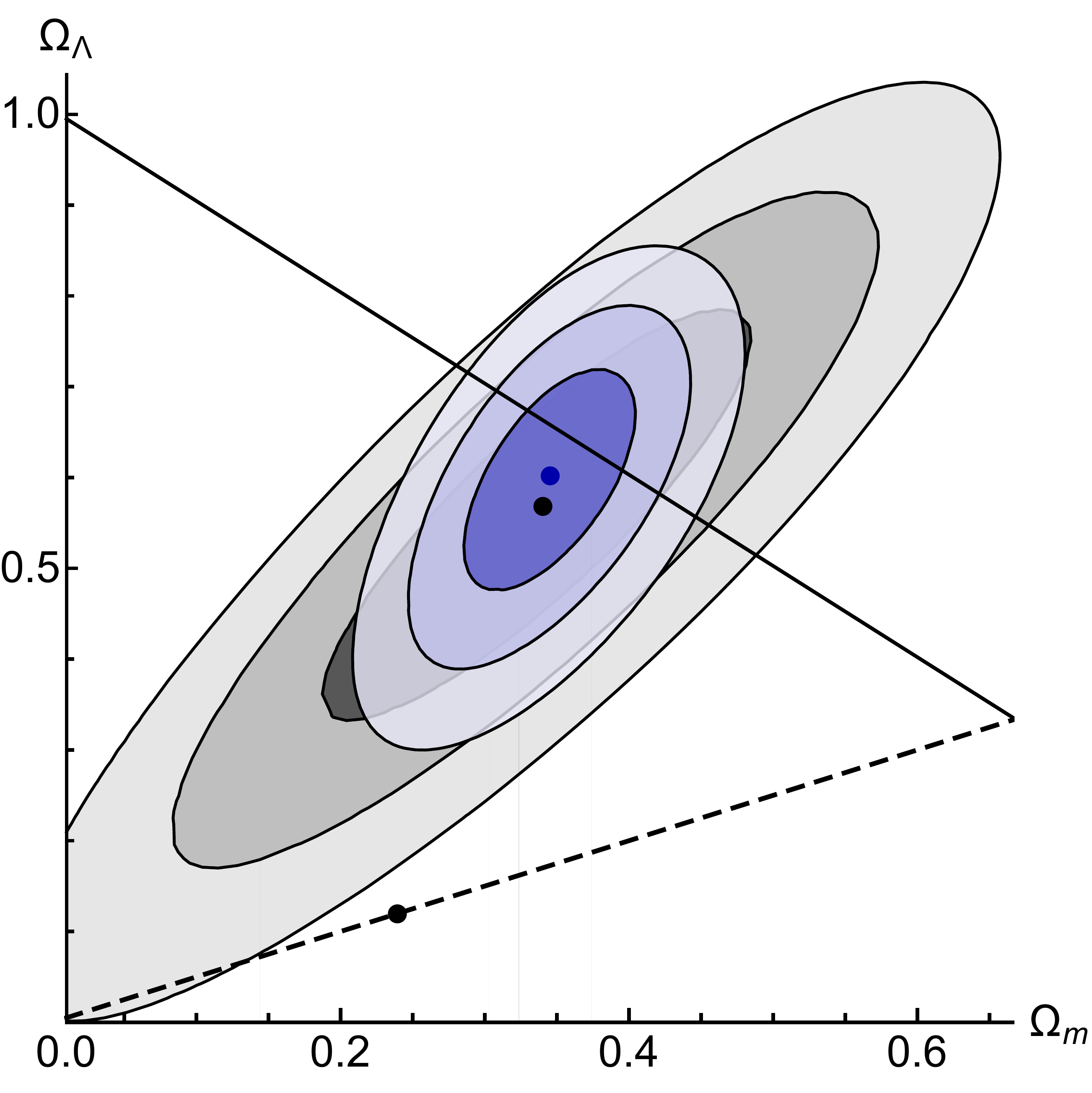}
\includegraphics[width=0.45\textwidth]{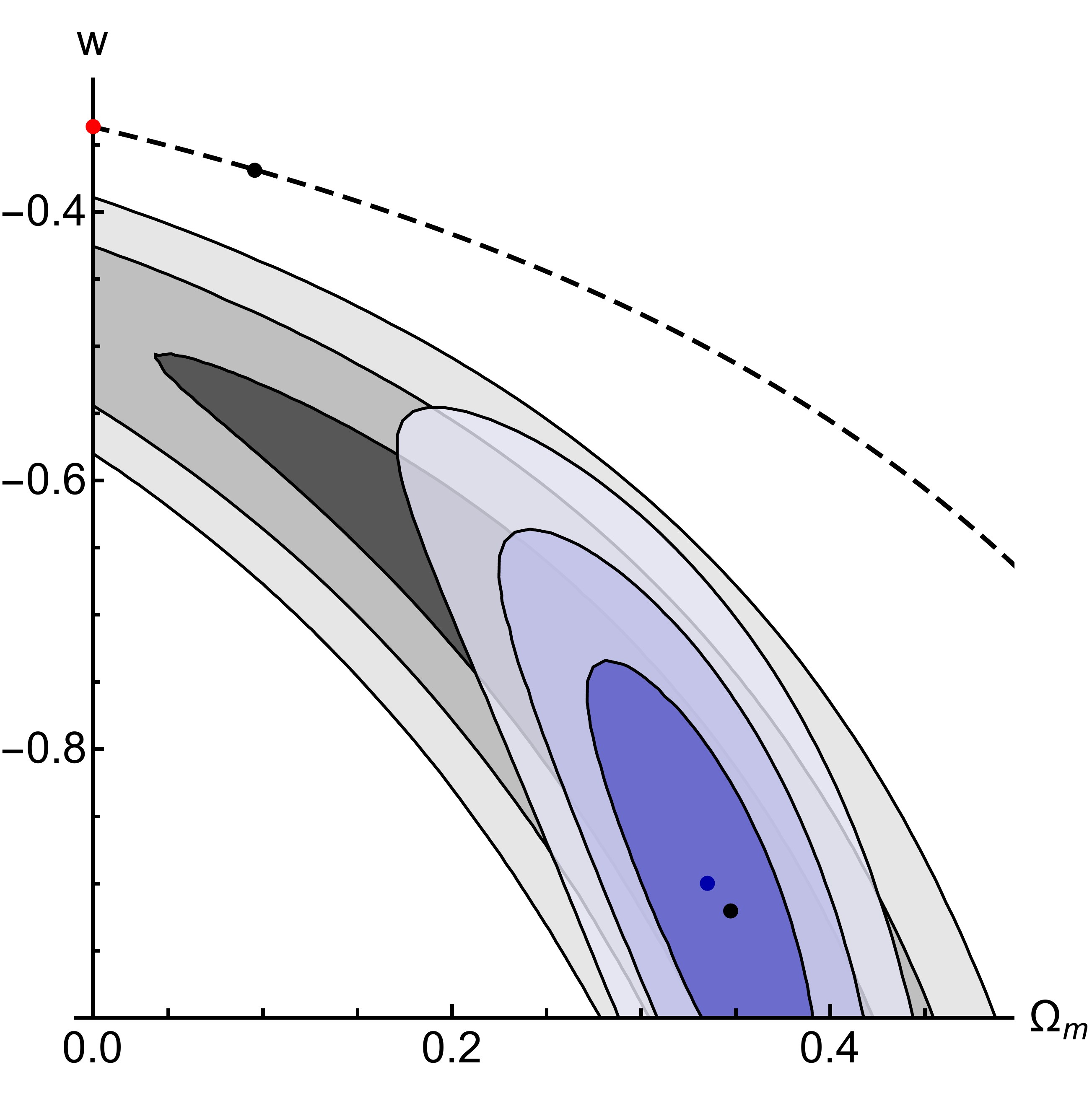}
\caption{The grey and violet confidence regions are obtained from the supernova alone and the joint analysis, respectively. Top panel: The 1, 2 and 3$\sigma$ confidence regions in the $ \Omega_{m} - \Omega_{\Lambda}$ parameter space for $k\Lambda$CDM. The solid line identifies the flat $\Lambda$CDM models. The dashed line gives the no-acceleration criterion: $\Omega_m = \Omega_{\Lambda}/2$. The black point is where the $4.98\sigma$ confidence contour of the joint likelihood touches the dashed line. Bottom panel: The 1,2 and 3$\sigma$ confidence regions in the $\Omega_{m} - w$ parameter space for flat $w$CDM. As in the top panel, the dashed curve identifies the no-acceleration criterion: $w = -1/(3(1-\Omega_m))$. Likewise, the black point is where the $5.38\sigma$ confidence contour of the joint likelihood touches the dashed line.}
\label{fig:con}
\end{figure}

The SN Ia Hubble diagram was claimed to be consistent with a uniform rate of expansion in \cite{Nielsen15} as the analysis in $k\Lambda$CDM model evades non accelerating criterion by only $\lesssim 3\sigma$. We reproduce this result and agree with this statement (see top panel of \Cref{fig:con}). However, there is a strong prejudice for a flat universe from the CMB data (\cite{Ade16}), and hence it is important to analyse SN Ia and other cosmological data in the context of a flat $w$CDM model. We find that the evidence for acceleration in the $\Omega_{m} - w$ plane is much more significant ($\geq 4.56\sigma$) compared to the marginal ($\geq 2.88\sigma$) found in the $\Omega_{m} - \Omega_{\Lambda}$ plane (see \Cref{fig:con}). The claimed marginal evidence for acceleration that corresponds to a very low matter density becomes more significant in both $k\Lambda$CDM and $w$CDM models when more physical values of $\Omega_m$ are considered. In \Cref{fig:con} contours for the best-fit regions of supernova dataset and our joint analysis are shown. The joint analysis improves the evidence for acceleration in the $k\Lambda$CDM model to $4.98\sigma$ and in $w$CDM model to $5.38\sigma$. 

The parameter space $\Omega_{m} - w$ allows us to explore points that correspond to the functional forms of $R_{h}=ct$ and power-law models for specific values of the parameters in $w$CDM. The point $(\Omega_{m},w)=(0,-1/3)$, shown in red in the bottom panel of \Cref{fig:con}, phenomenologically reproduces the expansion law of the $R_{h}=ct$ model (cf. \Cref{eqn:HUE}). Similarly, the point $(\Omega_{m},w)=(0,-0.38)$ corresponds to the best-fit value of $n=1.08$ in power-law model from our joint analysis (cf. \Cref{tab:res} and \Cref{eqn:wn}).  Using SN data alone, the best-fit model for the power-law cosmology is found to be $ n = 1.28$ (corresponding to $w = -0.48$), implying an accelerating universe (cf. \Cref{sec:mod}). In the $\Omega_{m} - w$ plane this point lies within the $2\sigma$ SN confidence region, in contrast to the $R_h=ct$ at $4.56\sigma$.  However, when the joint analysis is performed, these points are at $5\sigma$ and $5.5\sigma$ for the power-law and $R_h=ct$ models, respectively.

The best-fit parameter values for the models considered in this work are presented in \Cref{tab:res}. The values for the fitted $r_{d}$ parameter are consistent with the estimates (see \cite{Ade16}) using the fit function for the drag epoch from \cite{Eisenstein98}. In our analysis, the best-fit value for the index $n$ is driven towards unity. This result is quite different form the previous $n$ estimates \citep{Dolgov14,Shafer15,Rani15}, which consistently suggest $n \sim 1.5$. The modification in the statistical method for the SN analysis enables this change. While, the value we find for BAO data alone ($n=0.94$) is consistent with the value given by \cite{Shafer15} ($n=0.93$), for BAO+SN data we find $n=1.11$ in contrast to his $n=1.52$. Our best-fit value for the joint analysis ($n = 1.08 \pm 0.02 $) is now consistent with $ n = 1.14 \pm 0.05$ found by \cite{Zhu08} using X-ray cluster data.

It is clear that the $H_{0}$ estimates for the power-law and $R_{h}=ct$ models are highly in tension with the direct estimate in \cite{Riess16}, which is already a well established problem for $\Lambda$CDM \citep{Lukovic16, Bernal16}. In this work, the joint analysis provides $H_0=66.4\pm1.8$ km s$^{-1}$/Mpc for the $\Lambda$CDM scenario. We note that this value is consistent with our previous estimate \citep{Lukovic16} but with a higher error due to the difference in the BAO analysis (see \Cref{sec:data}). In any case, this value still remains in tension with the direct estimate at $2.7\sigma$.

\begin{table}[h]
\begin{center}
\caption{Best-fit parameters for the joint analysis of SN+BAO+DA+GRB datasets, with 1$\sigma$ errors are reported here. We do not quote the parameters of SN and GRB models which are considered as nuisance parameters.}
\label{tab:res}
\footnotesize 
\resizebox{.5\textwidth}{!}{
\begin{tabular}{|c|c|c|c|c|}
\hline 
Model &$H_{0}$[km s$^{-1}$/Mpc] & $n$ & $\Omega_{m}$&$ r_{d}$[Mpc]  \\
\hline
$R_{h}=ct$ & 62.4$\pm$1.4 &1. &- &148.3$\pm$3.6 \\
Power-law & 64.2$\pm $1.7& 1.08$\pm$0.04 &- & 147.0$\pm$3.6 \\
$\Lambda$CDM & 66.4$\pm$1.8&- & 0.361$\pm$0.023 & 148.6$\pm$3.7\\
\hline
\end{tabular}
}
\end{center}
\end{table}

We want to stress that the Milne model with $\Omega_{m}=\Omega_{\Lambda}=0$ and $\Omega_{k}=1$ does not correspond to the flat $R_{h}=ct$ model. In fact, these two models share the same Hubble equation and EoS ($\rho + 3 p = 0$), but do not have the same $D_{L}$ as the negative curvature in the Milne model corresponds to $D_{L}\propto (1+z) \sinh(\log(1+z))$ with $\Omega_{k}=1$, where as in the $R_{h}=ct$ model $D_{L}\propto (1+z) \log(1+z)$. In the SN Ia Hubble diagram it is difficult to see any significant difference among the $\Lambda$CDM, Milne and $R_{h}=ct$ model predictions, however, their performance can be more effectively tested with the information criteria. 

\begin{table}[h]
\caption{$\Delta$(AIC)  and $\Delta$(BIC) comparisons for models with $\Lambda$CDM as the reference. 'Joint' corresponds to the joint analysis with SN+BAO+DA+GRB datasets.}
\label{tab:aic}
\begin{center}
\footnotesize 
\begin{tabular}{|c|c|c|c|c|}
\hline 
& $\Delta\text{(AIC)}_{\text{Joint}}$& $\Delta\text{(AIC)}_{\text{SN}}$& $\Delta\text{(BIC)}_{\text{Joint}}$& $\Delta\text{(BIC)}_{\text{SN}}$ \\
\hline
Power-law & 28.02 & 2.0 & 28.02 & 2.0 \\
$R_{h}=ct$ & 30.83  & 21.79 & 26.05 & 17.20 \\
Milne & 66.39  & 9.78 & 61.62 & 5.19 \\
\hline
\end{tabular}
\end{center}
\end{table}

The analysis of SN Union 2.1 data done by \cite{Melia12b} calls for a non accelerating scenario, as the $R_{h}=ct$ model was claimed to perform on par with $\Lambda$CDM. This point was also taken by \cite{Nielsen15} who analysed the JLA dataset with their improved statistical method. In our work, using the same technique for analysing the JLA data we find that $R_{h}=ct$ performs poorly when compared to $\Lambda$CDM with $\Delta\text{(AIC)}_{\text{SN}} \sim 22$, while a power-law model is performing as good as $\Lambda$CDM with $n\sim1.28$ from SN data. Our results are consistent with the previous work by \cite{Shafer15}. The values of $\Delta\text{(AIC)}_{\text{SN}}$ obtained from the SN data alone are shown in \Cref{tab:aic}. 

Note that the Milne model was claimed to perform marginally worse in comparison to $\Lambda$CDM using the SN data alone \citep{Nielsen15}. In our work, we find for this model $\Delta\text{(AIC)}_{\text{SN}}= 9.8$ high enough to reject a model. In any case, the Milne model fails to keep up when the high redshift GRB ($\Delta\text{(AIC)}_{\text{GRB}}\sim20$) and BAO ($\Delta\text{(AIC)}_{\text{BAO}}\sim38$) data are used, yielding a total $\Delta\text{(AIC)}_{\text{Joint}}\sim66.4$ (see \Cref{tab:aic}). The AIC statistics disfavours the power law models by $\Delta(\text{(AIC)}_{\text{Joint}})\sim 28$ and the $R_{h}=ct$ model by $\Delta(\text{(AIC)}_{\text{Joint}})\sim 30$ in comparison to the $\Lambda$CDM model. In any case, our joint analysis shows that all the three models (Milne, power-law and $R_h=ct$) are strongly disfavoured with respect to $\Lambda$CDM (\Cref{tab:aic}). 

\section{Conclusions}
\label{sec:con}

Contrary to the claim by \cite{Nielsen15}, we find that the SN data alone indicates an accelerating universe at more than $4.56 \sigma$ confidence level. This evidence becomes even stronger ($5.38\sigma$), when we perform the joint analysis combining SN, BAO, OHD and GRB data. The non accelerating $R_{h}=ct$ model fails to explain at once these data resulting in $\Delta\text{(AIC)}_{\text{Joint}} \sim 30$ with respect to $\Lambda$CDM. Although, the power-law model performs slightly better that the $R_{h}=ct$ model, similarly fails with a $\Delta\text{(AIC)}_{\text{Joint}} \sim 28$. Our analysis shows that the possibility of having models with an uniform rate of expansion is excluded given the current low-redshift data. In conclusion, on one hand we re-assert that the current expansion of our universe is accelerated and on the other hand that $\Lambda$CDM still constitutes the base line for a concordance model in cosmology. 

\bibliographystyle{aa}
\bibliography{Rhct}
 
\end{document}